\documentclass[aps,pra,prabib,twocolumn,showpacs,nofootinbib]{revtex4}
\usepackage{graphicx} \usepackage{amsmath} \usepackage{amssymb}
\usepackage{amsfonts} \usepackage{bm}

\begin{document}

\newcommand{\be}{\begin{equation}} \newcommand{\ee}{\end{equation}}
\newcommand{\bea}{\begin{eqnarray}}\newcommand{\eea}{\end{eqnarray}}

\title{Noncommutative oscillator, symmetry and Landau problem}

\author{Pulak Ranjan Giri} \email{pulakranjan.giri@saha.ac.in}

\affiliation{Theory Division, Saha Institute of Nuclear Physics,
1/AF Bidhannagar, Calcutta 700064, India}

\author{P Roy} \email{pinaki@isical.ac.in}

\affiliation{Physics and applied mathematics Unit, Indian Statistical
Institute, kolkata 700108, India}

\begin{abstract}
Isotropic oscillator on a plane is discussed where both the
coordinate and momentum space are considered to be noncommutative.
We also discuss the symmetry properties of the oscillator for three
separate cases when both the noncommutative parameters $\Theta$ and
$\overline{\Theta}$ satisfy specific relations. We compare the
Landau problem with the isotropic oscillator on noncommutative space
and obtain a relation between the two noncommutative parameters with
the magnetic field of the Landau problem.
\end{abstract}


\date{\today}

\maketitle

\section{Introduction}
In recent years noncommutative geometry
\cite{michael,bala1,rban2,pedro1,olmo,horvathy,horv11,michal,vanheke,pedro}
has received much attention due to the fact that spacetime may be
noncommutative at very small length scale. For detail study on
noncommutative space see the list of Refs. of   Refs.
\cite{pulak,pulak1}. Although the noncommutative scale is expected
to be pretty small, perhaps below the Planck scale \cite{gieres},
people are looking for phenomenological consequences of the
noncommutative geometry in low energy quantum mechanical regime. One
such example is the fractional angular momentum quantum number and
the existence of the zero point expectation value of the angular
momentum in noncommutative space, which has been discussed in a
recent paper \cite{zhang}.

Usually only the coordinate space is considered to be noncommutative
when the noncommutative physics is studied.  But in this letter we
consider both the coordinate and momentum space to be
noncommutative. However momentum space non-commutativity is not a
new concept. As we know in quantum mechanics the generalized
momentum components are noncommutative, which can be taken as a
motivation for us to study the quantum mechanical model in
noncommutative momentum space. Also recently it has been shown in
Ref. \cite{zhang} that in order to keep the Bose Einstein statistics
intact at the noncommutative level the momentum space needs to be
noncommutative besides the noncommutative coordinate space.

This letter is organized as follows: In section
\ref{s1} we consider the isotropic $2D$ oscillator in a space where
both the coordinate and momentum are noncommutative. The spectrum of
the system is found and its symmetry is discussed for a specific
noncommutative parameters range.  In section \ref{s2} the model of
section \ref{s1} is compared with the Landau problem and a relation
between the noncommutative parameters $\Theta$ and $\overline{\Theta}$
with the magnetic field is established. Finally
section \ref{s3} is devoted to a discussion.

\section{Isotropic oscillator in noncommutative space}\label{s1}
We consider  both the position and the momentum space to be
noncommutative. The noncommutative algebra in the phase space can be
written as (in $\hbar=1$ unit)
\begin{eqnarray}
\nonumber\left[\overline{x_i},\overline{x_j}\right]&=&
2i\epsilon_{ij}\Theta,~
\left[\overline{p_i},\overline{p_j}\right]=2i\epsilon_{ij}\overline{\Theta},\\
\left[\overline{x_i},\overline{p_j}\right]&=&
i\delta_{ij}\left[1+\Theta\overline{\Theta}\right]\,, \label{al4}
\end{eqnarray}
where $\Theta$ is the noncommutative parameter for the coordinate
space, $\overline{\Theta}$ is the noncommutative parameter for the
momentum space and $\epsilon_{11}=\epsilon_{22}=0, \epsilon_{12}=
-\epsilon_{21}=1$. The fact that the noncommutative parameter 
$\overline{\Theta}$ is not a magnetic field will be discussed  
later in this paper. Basically one can see that our algebra (\ref{al4}) is
unaffected by the electric charge  $e$ of the quantum particle. But in case of
magnetic field the algebra should be affected by the limit $e=0$.
The 2D isotropic oscillator  Hamiltonian
\begin{eqnarray}
H = \frac{1}{2m}{\boldsymbol{p}^2}+
\frac{1}{2}m\omega^2{\boldsymbol{r}}^2\,, \label{2dh}
\end{eqnarray}
can now be defined as the same Hamiltonian but now the coordinate
and momentum are replaced by the corresponding noncommutative
counter parts. The noncommutative counterpart of the above
Hamiltonian is \cite{michal,pulak1}
\begin{eqnarray}
\overline{H} = \frac{1}{2m}{\overline{\boldsymbol{p}}^2}+
\frac{1}{2}m\omega^2{\overline{\boldsymbol{r}}}^2\,. \label{2dhn}
\end{eqnarray}
It is possible to go back to the commutative space by replacing the
noncommutative operators in terms of the commutative operators. We
therefore need to get a representation of the algebra (\ref{al4}) in terms of
the commutative coordinates.
One possible representation is given by
\begin{eqnarray}
\nonumber \overline{x_1}&=&x_1 -\Theta p_2,~~ \overline{x_2}=x_2 +\Theta p_1\,,\\
\overline{p_1}&=&p_1 +\overline{\Theta} x_2,~~\overline{p_2}=p_2
-\overline{\Theta} x_1\,,
 \label{rep3}
\end{eqnarray}
The above representation is consistent with the algebra (\ref{al4}) and
can be used to go back to commutative
space, where the problem can be handled easily. Our Hamiltonian
(\ref{2dh}) thus becomes
\begin{eqnarray}
H_{\Theta,\overline{\Theta}} =
\frac{1}{2M_\Theta}{\boldsymbol{p}^2}+
\frac{1}{2}M_\Theta\Omega_{\Theta,\overline{\Theta}}^2{\boldsymbol{r}}^2
- S_{\Theta,\overline{\Theta}}L_z\,, \label{2dh1}
\end{eqnarray}
where $1/M_{\Theta}= 1/m +m\omega^2\Theta^2$,
$\Omega_{\Theta,\overline{\Theta}}= \sqrt{\left(1/m
+m\omega^2\Theta^2\right)\left(m\omega^2
+\overline{\Theta}^2/m\right)}$ and
$S_{\Theta,\overline{\Theta}}=m\omega^2\Theta +\overline{\Theta}/m$.

This Hamiltonian can be solved exactly. It is possible to write the
Hamiltonian (\ref{2dh1}) as a sum of two separate 1D harmonic
oscillators with frequencies $\Omega_{\Theta,\overline{\Theta}}^+,
\Omega_{\Theta,\overline{\Theta}}^-$
\cite{nair,agni,Smailagic,pghosh},
\begin{eqnarray}
\nonumber H_{\Theta,\overline{\Theta}}=
\Omega_{\Theta,\overline{\Theta}}^+\left(a_{\Theta,\overline{\Theta}}^\dagger
a_{\Theta,\overline{\Theta}} +1/2\right)+\\
\Omega_{\Theta,\overline{\Theta}}^-\left(b_{\Theta,\overline{\Theta}}^\dagger
b_{\Theta,\overline{\Theta}} +1/2\right) \label{2dosh}
\end{eqnarray}
where the annihilation operators $a_{\Theta,\overline{\Theta}},
b_{\Theta,\overline{\Theta}}$,
\begin{eqnarray}
a_{\Theta,\overline{\Theta}}=\frac{1}{2\sqrt{M_\Theta\Omega_{\Theta,\overline{\Theta}}^+}}\left[(p_1+ip_2)
- iM_\Theta\Omega_{\Theta,\overline{\Theta}}^+(x_1+ix_2) \right]\\
b_{\Theta,\overline{\Theta}}=\frac{1}{2\sqrt{M_\Theta\Omega_{\Theta,\overline{\Theta}}^-}}\left[(p_1-ip_2)
- iM_\Theta\Omega_{\Theta,\overline{\Theta}}^-(x_1-ix_2) \right]
\label{annihi1}
\end{eqnarray}
and the corresponding creation operators
$a_{\Theta,\overline{\Theta}}^\dagger,
b_{\Theta,\overline{\Theta}}^\dagger$ satisfy the usual commutation
relations
\begin{eqnarray}
\left[a_{\Theta,\overline{\Theta}},
a_{\Theta,\overline{\Theta}}^\dagger\right]=\left[b_{\Theta,\overline{\Theta}},
b_{\Theta,\overline{\Theta}}^\dagger\right] =1\,,\label{2doshac}
\end{eqnarray}
with all other commutators being zero. The explicit form of the
frequencies are given by
\begin{eqnarray}
\Omega_{\Theta,\overline{\Theta}}^+=
\sqrt{\Omega_{\Theta,\overline{\Theta}}^2-S_{\Theta,\overline{\Theta}}^2}
+S_{\Theta,\overline{\Theta}}\,,\\
\Omega_{\Theta,\overline{\Theta}}^-=
\sqrt{\Omega_{\Theta,\overline{\Theta}}^2-S_{\Theta,\overline{\Theta}}^2}
-S_{\Theta,\overline{\Theta}}\,, \label{omega1}
\end{eqnarray}
where $\Omega_{\Theta,\overline{\Theta}}^+ \neq
\Omega_{\Theta,\overline{\Theta}}^-$~. The number operators
$\mathcal{N}_{\Theta,\overline{\Theta}}^+=a_{\Theta,\overline{\Theta}}^\dagger
a_{\Theta,\overline{\Theta}}$ and
$\mathcal{N}_{\Theta,\overline{\Theta}}^-=b_{\Theta,\overline{\Theta}}^\dagger
b_{\Theta,\overline{\Theta}}$ satisfy the eigenvalue equation
$\mathcal{N}_{\Theta,\overline{\Theta}}^\pm|n^+,n^-\rangle =
n^\pm|n^+,n^-\rangle$ with $n^\pm= 0,1,2,3,....$. Now the exact
eigenvalue for the Hamiltonian (\ref{2dosh}) is known in literature
and is of the form
\begin{eqnarray}
E_{\Theta,\overline{\Theta}}=
\Omega_{\Theta,\overline{\Theta}}^+(n^++1/2)
+\Omega_{\Theta,\overline{\Theta}}^-(n^-+1/2)\,, \label{2dosei}
\end{eqnarray}
which is  non-degenerate due to the presence of non-commutativity.
But it is also possible to get back the $su(2)$ symmetry for a
specific case when  the ratio of the two frequencies is rational,
$\Omega_{\Theta,\overline{\Theta}}^+/\Omega_
{\Theta,\overline{\Theta}}^-=a^-/a^+$, $a_-, a_+$ are relatively prime numbers.
But before going into the complicated case we first mention the
simplest case, $\overline{\Theta} =- m^2\omega^2\Theta$, when  also
the $su(2)$ symmetry is recovered. Since  for $\overline{\Theta} =-
m^2\omega^2\Theta$ the third term in the Hamiltonian (\ref{2dh1}) is
zero the Hamiltonian becomes isotropic.  So the the eigenvalue becomes
\begin{eqnarray}
E_{\Theta,\overline{\Theta}=-m^2\omega^2\Theta}= \left(n^+ + n^-
+1\right)\Omega_\Theta\,, \label{2dh1e}
\end{eqnarray}
where the frequency is $\Omega_\Theta=
\omega\left(1+m^2\omega^2\Theta^2\right)=
\omega\left(1-\Theta\overline{\Theta}\right)$.  Another important
case is when
\begin{eqnarray}
\Omega_{\Theta,\overline{\Theta}}=
S_{\Theta,\overline{\Theta}}\Rightarrow
\Theta\overline{\Theta}=1\,.\label{cons1}
\end{eqnarray}
Due to the constraint (\ref{cons1}) the Hamiltonian is infinitely
degenerate and the eigenvalues are given by
\begin{eqnarray}
E_{\Theta\overline{\Theta}=1}=
\left(2n+1\right)\widehat{\Omega_\Theta}\,, \label{comdeei}
\end{eqnarray}
where $\widehat{\Omega_\Theta}= m\omega^2\Theta +\frac{1}{\Theta m}$
and $n=0,1,2,..$. Note that the above spectrum is obtained in
Ref. \cite{nair}, but there the origin of momentum noncommutativity is due to
magnetic field and thus coupled to the electric charge of the particle. 
Now let us consider the non degenerate spectrum
(\ref{2dosei}) which is known \cite{louck} to have $su(2)$ symmetry
in a certain case. Let us assume that for $n^\pm= a^\pm p^\pm +
b^\pm$, the ratio of the two frequencies be such that ,
$\Omega_{\Theta,\overline{\Theta}}^+
a^+=\Omega_{\Theta,\overline{\Theta}}^-
a^-=\widehat{\Omega_{\Theta,\overline{\Theta}}}$ for certain
constants $a^\pm$, $b^\pm$ and $p^\pm= 0,1,2,....$. Then the
spectrum (\ref{2dosei}) becomes
\begin{eqnarray}
E_{\Theta,\overline{\Theta}}=
\widehat{\Omega_{\Theta,\overline{\Theta}}}(p^+ + p^-) +
\Omega_{\Theta,\overline{\Theta}}^+ b^+ +
\Omega_{\Theta,\overline{\Theta}}^- b^- \,,\label{2doseisu2}
\end{eqnarray}
which has $su(2)$ symmetry. Note that the $\overline{\Theta}\to 0$
limit of (\ref{2doseisu2}) should reduce to the results already
obtained in Ref. \cite{agni}.
\begin{figure}
%\begin{center}
\includegraphics[width=0.4\textwidth, height=0.2\textheight]{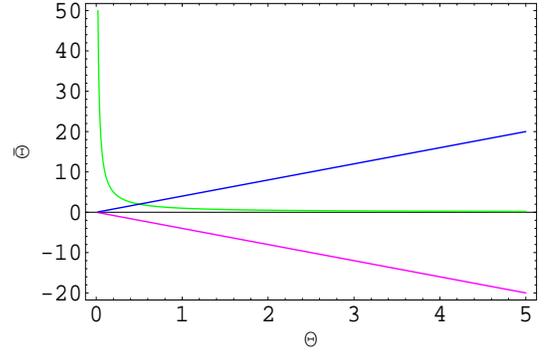}
\caption {(color online) A plot of $\overline{\Theta}$ as a function
of $\Theta$ for $m=1$, $\omega=2$. The green curve corresponds to
$\Theta\overline{\Theta}=1$ (it corresponds to a spectrum given in
Eq. (\ref{comdeei}), which is  infinitely degenerate), the bule
curve corresponds to $\overline\Theta= m^2\omega^2\Theta$ (it
restores BE statistics) and the pink curve corresponds to
$\overline\Theta= -m^2\omega^2\Theta$ (it corresponds to a spectrum
of Eq. (\ref{2dh1e}), which has $su(2)$ symmetry)}
%\end{center}
\end{figure}

\section{Landau problem and noncommutative space}\label{s2}

We now consider the Landau problem on a plane.  For different discussions on
Landau problems see Refs. \cite{horvathy,martin} The Hamiltonian in
symmetric gauge  $\boldsymbol{A}= B/2(-x_2, x_1)$ is given by
\begin{eqnarray}
H_B = \frac{1}{2\widetilde{m}}{\boldsymbol{p}^2}+
\frac{B^2}{8\widetilde{m}}{\boldsymbol{r}}^2
-\frac{B}{2\widetilde{m}}L_z\,, \label{2dhL}
\end{eqnarray}
where $B$ is the constant magnetic field in the $z$ direction. The
eigenvalues known as Landau levels  are known to be infinitely
degenerate
\begin{eqnarray}
E_{B}= (2n +1)\Omega_{B} \,, \label{2doseiB}
\end{eqnarray}
where $\Omega_{B}= \frac{B}{2\widetilde{m}}$. Note the similarity
between the Landau levels (\ref{2doseiB}) and (\ref{comdeei}). The
two spectrum will overlap with each other if
\begin{eqnarray}
\widehat{\Omega_\Theta}= \Omega_{B}\,, \mbox{for}
~~\Theta\overline{\Theta}=1\,. \label{fre2}
\end{eqnarray}
Due to this similarity between the two spectrum one can exploit it
and get a relation between the noncommutative parameters and the
magnetic field in the Landau problem. On comparing the mass term
of the two Hamiltonians, namely (\ref{2dh1}) and (\ref{2dhL}), we obtain
\begin{eqnarray}
M_\Theta= \widetilde{m}\,. \label{mass1}
\end{eqnarray}
Considering (\ref{fre2}) and (\ref{mass1}) together we get
\begin{eqnarray}
\overline{\Theta}=1/\Theta= B/2\,. \label{relation1}
\end{eqnarray}
We used the unit where $\hbar=c=e=1$. For the quantum hall experiment the
typical strength of  magnetic field is $B \sim 12T - 15T$. Then the
noncommutative parameters are typically
\begin{eqnarray}
\Theta= 1/{\overline\Theta} \sim 0.22\times 10^{-11}cm -
0.176\times 10^{-11}cm\,.\label{neumerical}
\end{eqnarray}
Note that by comparing term by term of the Hamiltonian of the Landau
problem and the Hamiltonian of an oscillator in noncommutative
coordinate space in Ref. \cite{gam} an inverse relation  between the
noncommutative parameter and magnetic field has been established,
which is comparable with the second and third term of Eq.
(\ref{relation1}) and numerical value of the noncommutative
parameter due to coordinate non-commutativity is also obtained. One
should keep in mind that the above relation (\ref{relation1}) is
just an analogy but in general there is now relation of the two
noncommutative parameters $\Theta$ and $\overline\Theta$ with the
magnetic field. Although from the representation of noncommutative
momentum given in Eq. (\ref{rep3}) one may think that it is analog
to the generalized momentum $\boldsymbol{P}=\boldsymbol{p}-
e\boldsymbol{A}$ in constant magnetic filed background written in
symmetric gauge for the vector potential $\boldsymbol{A}$ but this
is clearly not so. The reason they are different will be evident if one
consider neutral particle for which the noncommutative momentum
still satisfy the commutative relation (\ref{al4}) but in magnetic
field background now the generalized momentum
$\boldsymbol{P}=\boldsymbol{p}$ (because $e=0$), which means they
commute. In fact the momentum noncmmutativity is used for the
neutral particle model of neutron in a quantum gravitational well
\cite{bertolami} and bounds  for the noncommutative parameters have
been obtained using the experimental result \cite{nes1} of neutron
gravitational states.

\begin{figure}
%\begin{center}
\includegraphics[width=0.4\textwidth, height=0.2\textheight]{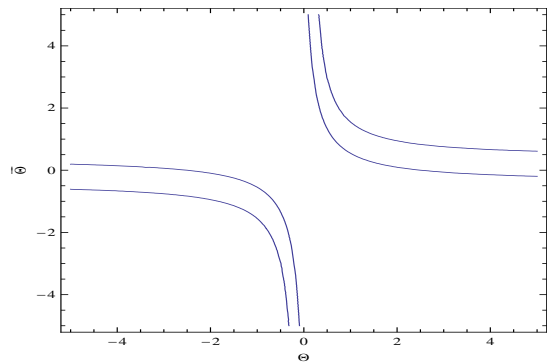}
\caption {(color online) A contour plot of
  $\Omega_{\Theta,\overline{\Theta}}^+/\Omega_{\Theta,\overline{\Theta}}^-=
  a^+/a_-$ for $a_+=2$, $a_-=3$, $m=1$, $\omega=2$ is drawn. Along the curve
  the $su(2)$ symmetry is restored. The spectrum along the curve  is
given by
  Eq. (\ref{2doseisu2}) with $a_+=2$, $a_-=3$, $m=1$, $\omega=2$}
\end{figure}

\section{Conclusion and Discussions}\label{s3}
Isotropic oscillator on a plane is considered in this article,
where we take both the
coordinate and momentum space to be noncommutative. We discussed its
general solution which is non degenerate. But it is possible to get
back the degenerate spectrum when the
noncommutative parameters $\Theta$ and $\overline{\Theta}$ satisfy certain constrants. For
example for $\overline{\Theta}= -m^2\omega^2\Theta$ the system
is $su(2)$ symmetric, the spectrum being given by Eq. (\ref{2dh1e}). Note that  for the
condition $\overline{\Theta}= -m^2\omega^2\Theta$
Bose-Einstein statistics does not remain intact at the
noncommutative level \cite{zhang}, which can be understood as
follows. On noncommutative space defined by the algebra (\ref{al4})
one can construct the deformed creation and annihilation operators
$\overline{a_i}=
\sqrt{\frac{1}{2m\omega}}\left(m\omega\overline{x_i} +
i\overline{p_i}\right)$ and $\overline{a_i}^\dagger=
\sqrt{\frac{1}{2m\omega}}\left(m\omega\overline{x_i} -
i\overline{p_i}\right)$ respectively  replacing the corresponding
commutative coordinate and momentum by its corresponding
noncommutative counterpart. For the two creation operators
$\left[\overline{a_1}^\dagger,\overline{a_2}^\dagger\right]\neq 0$,
for $\overline{\Theta}= -m^2\omega^2\Theta$, which indicates that BE statistics
is violated. But it is also possible
to keep Bose-Einstein statistics intact i.e.,
$\left[\overline{a_1}^\dagger,\overline{a_2}^\dagger\right]= 0$,
which will lead to $\overline{\Theta}= m^2\omega^2\Theta$. It is
known  from Ref. \cite{zhang} that the angular momentum
$\overline{L}= \overline{x_1}~ \overline{p_2}-\overline{x_2}~
\overline{p_1}$ does possess zero point eigenvalue unlike
commutative case where $\langle \mbox{g.s}|L=
x_1p_2-x_2p_1|\mbox{g.s}\rangle =0$. In our notation
$\overline{L}=\left(1+\Theta\overline{\Theta}\right)L-
m^2\omega^2\Theta\left(x_1^2+x_2^2\right)-
\Theta\left(p_1^2+p_2^2\right)$, for $\overline{\Theta}=
m^2\omega^2\Theta$. The zero point eigenvalue $\langle
\mbox{g.s}|\overline{L}|\mbox{g.s}\rangle =2m\omega\Theta$. The most
general algebra satisfied by $\overline{a_i}$,
$\overline{a_i}^\dagger$ is
$\left[\overline{a_i},\overline{a_i}^\dagger\right]=
\frac{1}{2m\omega}\left[2i(m^2\omega^2\Theta+\overline{\Theta})\epsilon_{ij}
+m\omega[1+\Theta\overline{\Theta}](\delta_{ij}+\delta_{ji})\right]$,~\\
$\left[\overline{a_i},\overline{a_i}\right]=\left[\overline{a_i}^\dagger,\overline{a_i}^\dagger\right]=
\frac{1}{2m\omega}\left[2i(m^2\omega^2\Theta-\overline{\Theta})\epsilon_{ij}\right]$.
It is to be noted from (\ref{al4}) that the Planck constant is
modified like $\overline{\hbar}= \left(1 +
\Theta\overline{\Theta}\right)\hbar$ due to simultaneous
consideration the coordinate and momentum space
noncommutativity as pointed out in  Ref. \cite{bertolami} and in other papers
also. Finally, it may be noted that instead of starting with an isotropic
oscillator on the noncommutative plane, one could have started with an
anisotropic oscillator on the noncommuting plane \cite{jellal}. In that case
only the mass $M_{\Theta,\overline\Theta}$ would change and the relation
concerning $\Theta$,
$\overline\Theta$ and $B$ will remain unchanged.


\begin{thebibliography}{99}
\bibitem{michael} M. R. Douglas and N. A. Nekrasov, Rev. Mod. Phys.
{\bf 73}, 977 (2001).


\bibitem{bala1} A. P. Balachandran, T. R. Govindarajan, C. Molina and
P. Teotonio-Sobrinho, JHEP 0410 (2004) 072.

\bibitem{rban2} R. Banerjee and K. Kumar, Phys. Rev. {\bf D75},
045008 (2007).

\bibitem{pedro1} P. D. Alvarez, J. Gomis, K. Kamimura and
M. S. Plyushchay, Annals Phys. {\bf 322} 1556 (2007).

\bibitem{olmo}   M. A. del Olmo and  M. S. Plyushchay,
Annals Phys, {\bf 321}, 2830 (2006).


\bibitem{horvathy} P. A. Horvathy, Annal. Phys. {\bf 299}, 128 (2002).

\bibitem{horv11}  P. A. Horvathy and  M. S. Plyushchay,
Nucl. Phys. {\bf B714}, 269 (2005).

\bibitem{michal}   M. Demetrian and , D. Kochan, Acta Physica Slovaca
{\bf 52}, 1 (2002).

\bibitem{vanheke}  F. J. Vanhecke,  C. Sigaud and  A.R.da Silva,
  Braz. J. Phys. {\bf 36}, 194 (2006).

\bibitem{pedro} P. D. Alvarez, J. Gomis, K. Kamimura and
M. S. Plyushchay, Phys. Lett. {\bf B659}, 906 (2008).



\bibitem{pulak} P. R. Giri, arXiv:0801.0356v1 [hep-th].

\bibitem{pulak1} P. R. Giri, arXiv:0802.05516v2 [hep-th].

\bibitem{gieres} F. Delduc, Q. Duret, F. Gieres and M. Lefrancois,
arXiv:0710.2239v1 [quant-ph].



\bibitem{zhang} Jian-zu Zhang, Phys. Lett. {\bf B584}, 204 (2004).

\bibitem{agni} A. Kijanka and P. Kosinski, Phys. Rev. {\bf
D70}, 127702 (2004).

\bibitem{nair}  V.P. Nair and A.P. Polychronakos, Phys.Lett. {\bf B505}, 267
(2001).

\bibitem{Smailagic} A. Smailagic and E. Spallucci, Phys. Rev. {\bf D65}, 107701 (2002).


\bibitem{pghosh} P. K. Ghosh, Eur. Phys. J {\bf C42}, 355 (2005).


\bibitem{louck} J. D. Louck, M. Moshinsky and K. B. Wolf, J. Math.
Phys. {\bf 14}, 692 (1973).


\bibitem{martin} A. Bermudez, M.A. Martin-Delgado, E. Solano
Phys. Rev. Lett. {\bf 99}, 123602 (2007)


\bibitem{gam} J. Gamboa, M. Loewe, F. Mendez and J. C. Rojas, Mod.
Phys. Lett. {\bf A16}, 2075 (2001).


\bibitem{bertolami} O. Bertolami, J. G. Rosa, C. M. L. de Aragao, P.
Castorina and D. Zappala, Phys. Rev. {\bf D72}, 025010 (2005).



\bibitem{nes1} V. V. Nesvizhevsky et al.,  Nature {\bf 415}, 297 (2002).

\bibitem{jellal} A. Jellal, E.H. El Kinani and M. Schreiber, Int.J.Mod.Phys
  {\bf A20}, 1515 (2005).




\end{thebibliography}
\end{document}